\documentclass[aps, prd, 10pt, twocolumn, superscriptaddress, noshowpacs, preprintnumbers, longbibliography, groupedaddress, footinbib, bibnotes]{revtex4-1}

\usepackage{amsmath}
\usepackage{amsfonts}
\usepackage{amssymb}
\usepackage{bbold}
\usepackage{epsfig}
\usepackage{graphicx}
\usepackage{bm}
\usepackage{array}
\usepackage{hyperref}
\usepackage{listings}
\usepackage{color}
\usepackage{float}
\usepackage[normalem]{ulem}

\begin{document}

\title{Symmetry breaking induced by pairwise conversion of neutrinos in compact  sources}

\author{Shashank Shalgar}
\email{shashank.shalgar@nbi.ku.dk}
\thanks{ORCID: \href{http://orcid.org/0000-0002-2937-6525}{0000-0002-2937-6525}}
\affiliation{Niels Bohr International Academy \& DARK, Niels Bohr Institute,\\University of Copenhagen, Blegdamsvej 17, 2100 Copenhagen, Denmark}

\author{Irene Tamborra}
\email{tamborra@nbi.ku.dk}
\thanks{ORCID: \href{http://orcid.org/0000-0001-7449-104X}{0000-0001-7449-104X}}
\affiliation{Niels Bohr International Academy \& DARK, Niels Bohr Institute,\\University of Copenhagen, Blegdamsvej 17, 2100 Copenhagen, Denmark}

\date{\today}

\begin{abstract}
A surprising consequence of  non-linear flavor evolution  is the spontaneous breaking of the initial symmetries of the neutrino gas propagating in a dense astrophysical environment. 
 We explore the flavor conversion physics  by taking into account the polar and azimuthal angular distributions of neutrinos and present the very first example of spontaneous symmetry breaking in the context of fast flavor mixing in the nonlinear regime. Intriguingly, we find that fast flavor mixing does not always develop in the proximity of  the angular regions with vanishing  electron lepton number, as commonly assumed, and large flavor mixing can rapidly spread through all neutrino modes.
Such behavior  cannot be predicted from the linear regime of the flavor evolution. These results  can have major consequences on the physics of compact astrophysical objects.

\end{abstract}

\maketitle

\section{Introduction}
\label{sec:intro}

In neutrino dense environments, the  flavor evolution of neutrinos is affected by the interaction of neutrinos among themselves~\cite{Duan:2010bg,Mirizzi:2015eza,Chakraborty:2016yeg,Tamborra:2020cul}. 
 The non-linear feedback in the neutrino equations of motion is too complicated to be understood analytically, and not enough progress has been made on the numerical front.
A recent development concerns the possibility that fast pairwise conversion occurs~\cite{Sawyer:2015dsa,Sawyer:2008zs,Sawyer:2005jk, Chakraborty:2016yeg,Tamborra:2020cul}.
In the case of ordinary interactions of neutrinos with matter, leading to the well known Mikheev-Smirnov-Wolfenstein (MSW) effect~\cite{Mikheev:1986if,1985YaFiz..42.1441M,1978PhRvD..17.2369W}, flavor mixing depends on the  energy of (anti)neutrinos. Fast pairwise conversion is instead driven by the angular distributions of (anti)neutrinos~\cite{Sawyer:2015dsa,Sawyer:2008zs,Sawyer:2005jk,Chakraborty:2016lct,Izaguirre:2016gsx,Airen:2018nvp}; however, the dependence of neutrino flavor evolution on neutrino energy cannot be ignored in the non-linear phase~\cite{Shalgar:2020xns,Shalgar:2021wlj}.

The angular distribution of (anti)neutrinos is characterized by two independent angular variables, polar and azimuthal. 
Up to now, for the sake of simplicity, the azimuthal distribution of (anti)neutrinos has been neglected in most cases under the assumption of azimuthal symmetry; we refer the interested reader to,  e.g., Refs.~\cite{DelfanAzari:2019epo,2019ApJ...886..139N,Chakraborty:2016lct} for preliminary work incorporating the azimuthal variable in the linear phase of neutrino pairwise conversion.
In the azimuthally symmetric framework,  it has been shown that  fast flavor instabilities develop in the  proximity of crossings in the polar angular distribution of the electron neutrino lepton number (ELN)~\cite{Izaguirre:2016gsx,Dasgupta:2016dbv,Yi:2019hrp,Martin:2019gxb,Tamborra:2020cul,Morinaga:2021vmc}. The existence of such flavor instability  depends on the steepness and  the depth of the ELN crossing~\cite{Yi:2019hrp,Martin:2019gxb} and may be affected by collisions with the medium~\cite{Shalgar:2020wcx,Capozzi:2018clo,Shalgar:2019kzy,Johns:2021qby,Martin:2021xyl}. However, despite the existence of ELN crossings, significant fast flavor mixing may not be achieved~\cite{Padilla-Gay:2020uxa}, and the ELN crossings could  be  affected by neutrino advection dynamically~\cite{Shalgar:2019qwg}.

In this paper, we focus on angular distributions that are not azimuthally symmetric and investigate the flavor conversion physics. 
Azimuthal angular distributions could also be instrumental  in determining the existence of spontaneous symmetry breaking. The latter is a peculiar feature of collective flavor evolution discovered in the context of slow neutrino self-interaction~\cite{Raffelt:2013rqa, Duan:2014gfa, Abbar:2015mca}, but never  explored for   fast flavor conversion.

We carry out  numerical simulations of the fast flavor evolution in three flavors, including polar and azimuthal angular modes. For the first time, we show  that spontaneous symmetry breaking may significantly impact the fast flavor conversion rate in the non-linear regime.
Moreover, while    fast flavor evolution in the non-linear regime in an azimuthally symmetric system seems to imply that flavor conversion develops in the vicinity of the ELN crossings~\cite{Tamborra:2020cul}, this correlation is no longer valid for certain configurations of the zenith and azimuthal angular modes. Hence, contrary to common assumptions in the literature,  the correlation between the location of the ELN crossing and the onset of flavor evolution is a special case  observed in azimuthally symmetric configurations.

\section{Neutrino angular distributions}
\label{sec:angular}
At each point in space and time, a homogeneous neutrino  field can be expressed in terms of angular distributions that are functions of the polar and azimuthal angles, $\theta$ and 
$\phi$.  Assuming that the (anti)neutrinos are mono-energetic, for each $(\theta, \phi)$, the flavor information can be encoded in a $3 \times 3$ density matrix for (anti)neutrinos, $\rho$ ($\bar\rho$), whose  diagonal elements denote the occupation number of (anti)neutrinos for each flavor. 

When only the angular distribution in $\theta$ is considered, a necessary condition for  fast flavor instabilities is the occurrence of at least one ELN crossing~\cite{Izaguirre:2016gsx,Morinaga:2021vmc,Capozzi:2019lso}. 
When the azimuthal angular distribution is also taken into account, the ELN crossing occurs for a locus of points which can be plotted on a two dimensional surface, e.g.~using the Mollweide projection. We  classify the angular distribution through the number of disconnected {\it rings with vanishing ELN} (instead of the number of {\it ELN crossings} usually adopted in  azimuthally symmetric systems); e.g., an azimuthally symmetric distribution with one crossing would thus correspond to one ring of vanishing ELN, two crossings would correspond to two ELN rings, and so on.

To investigate the qualitative differences arising  in the flavor conversion physics,  we consider a set of representative configurations of the angular distributions for $\nu_e$'s and $\bar\nu_e$'s (Cases A, B, C, and D), two  with one ELN ring and two with  two ELN rings, respectively:
\begin{widetext}
\begin{eqnarray}
\label{all}
 {\rho}_{ee, \mathrm{A, B, C, D}} &=& \frac{1}{4\pi}\ ,\\
\label{casea}
\bar{\rho}_{ee, \mathrm{A}} &=& \frac{1}{2\pi}\left\{0.47 + 0.05 \exp\left[-(\cos\theta-1)^{2}\right]\right\}\ ,\\
\label{caseb}
\bar{\rho}_{ee, \mathrm{B}} &=& \frac{1}{2\pi}\left\{0.47 + 0.05 \left[\exp\left(-2 (\theta-\pi/2)^2\right)  \exp\left(-0.5 (\phi-\pi)^{2}\right)\right]\right\}\ ,\\
\label{cased}
\bar{\rho}_{ee, \mathrm{C}} &=& \frac{1}{2\pi}\left\{0.48 + 0.05 \left[\exp(-2.5 (\theta-\pi/2)^2)  \exp(-2 (\phi-\pi)^{2}) \right.\right. 
\left. \left. +\exp(-2 (\theta-\pi/2)^2) \exp(-2 (\phi-\pi)^{2}) \right]\right\}\ ,\\
\label{casee}
\bar{\rho}_{ee, \mathrm{D}} &=& \frac{1}{2\pi}\left\{0.49 + 0.05 \left[\exp(-2.5 (\theta-\pi/2)^2)  \exp(-2 (\phi-\pi)^{2}) \right.\right. 
\left. \left. +\exp(-2 (\theta-\pi/2)^2) \exp(-2 (\phi-\pi)^{2}) \right]\right\}.
\end{eqnarray}
\end{widetext}
We assume that the non-electron type neutrinos are initially negligible and are generated through flavor mixing.

In order to quantify the differences among  the angular distributions of Cases A--D, we introduce the parameters:
\begin{eqnarray}
\textrm{ELN$\uparrow$} &=& \begin{cases}
\frac{\int (\rho_{ee} - \bar{\rho}_{ee}) d\cos\theta d\phi}{\int (\rho_{ee} + \bar{\rho}_{ee}) d\cos\theta d\phi} & \mathrm{for}~\rho_{ee} \ge \bar{\rho}_{ee}\\
0  & \mathrm{otherwise} \ ,
\end{cases} 
\label{ELNup}
\\
\textrm{ELN$\downarrow$} &=& \begin{cases}
\frac{\int (\bar{\rho}_{ee} - \rho_{ee}) d\cos\theta d\phi}{\int (\rho_{ee} + \bar{\rho}_{ee})  d\cos\theta d\phi}& \mathrm{for}~\rho_{ee} \le  \bar{\rho}_{ee}\\  
0  & \mathrm{otherwise} \ .
\end{cases} 
\label{ELNdown}
\end{eqnarray}
The parameters $\textrm{ELN$\uparrow$}$ and $\textrm{ELN$\downarrow$}$ help to  distinguish among configurations with similar total ELN, but overall different angular distributions. For Cases A--D, the values of ELN$\uparrow$ and ELN$\downarrow$ are reported in Table~\ref{Tab1} to illustrate the relative excess of $\rho_{ee}$ or $\bar{\rho}_{ee}$ in the two regions  separated by vanishing ELN, together with 
 the difference between ELN$\uparrow$ and ELN$\downarrow$ which gives the overall excess of $\nu_e$  over $\bar\nu_e$ normalized  to the total number of  $\nu_e$  and $\bar\nu_e$.
\begin{table}[b]
\caption{Parameters characterizing the initial angular distributions for Cases A--D. The second and third columns represent  ELN$\uparrow$ and ELN$\downarrow$ (Eqs.~\ref{ELNup} and \ref{ELNdown}),  respectively. The fourth column is the difference between the first two columns and is indicative of the commonly adopted total ELN number.  The last column tabulates the number of ELN rings.
}
\begin{tabular}{|l|l|l|l|l|}
\hline
 & ELN$\uparrow$ & ELN$\downarrow$ & ELN$\uparrow$-ELN$\downarrow$ & \# of rings \cr
\hline
\hline
Case A & 0.01257 & 0.00456 & 0.00801 & 1\cr
\hline
Case B & 0.00527 & 0.00529 & -0.00002 & 1\cr
\hline
Case C & 0.01254 & 0.00256 & 0.00998 & 2\cr
\hline
Case D & 0.00499 & 0.00510 & -0.00012 & 2 \cr
\hline
\end{tabular}
\label{Tab1}
\end{table}
From Table~\ref{Tab1}, we can clearly see that Cases A and C have an excess of   $\nu_e$ over $\bar\nu_e$, while  Cases B and D  have a very small total ELN number (with respect to Cases A and C) and an excess of $\bar\nu_e$ over $\nu_e$. It is worth highlighting that Cases B and D do not have  similar distributions of $\nu_e$ and $\bar\nu_e$, although they both have a small total ELN, as shown by the values of $\textrm{ELN$\uparrow$}$ and $\textrm{ELN$\downarrow$}$ in Table~\ref{Tab1}.

\begin{figure*}
\includegraphics[width=0.49\textwidth]{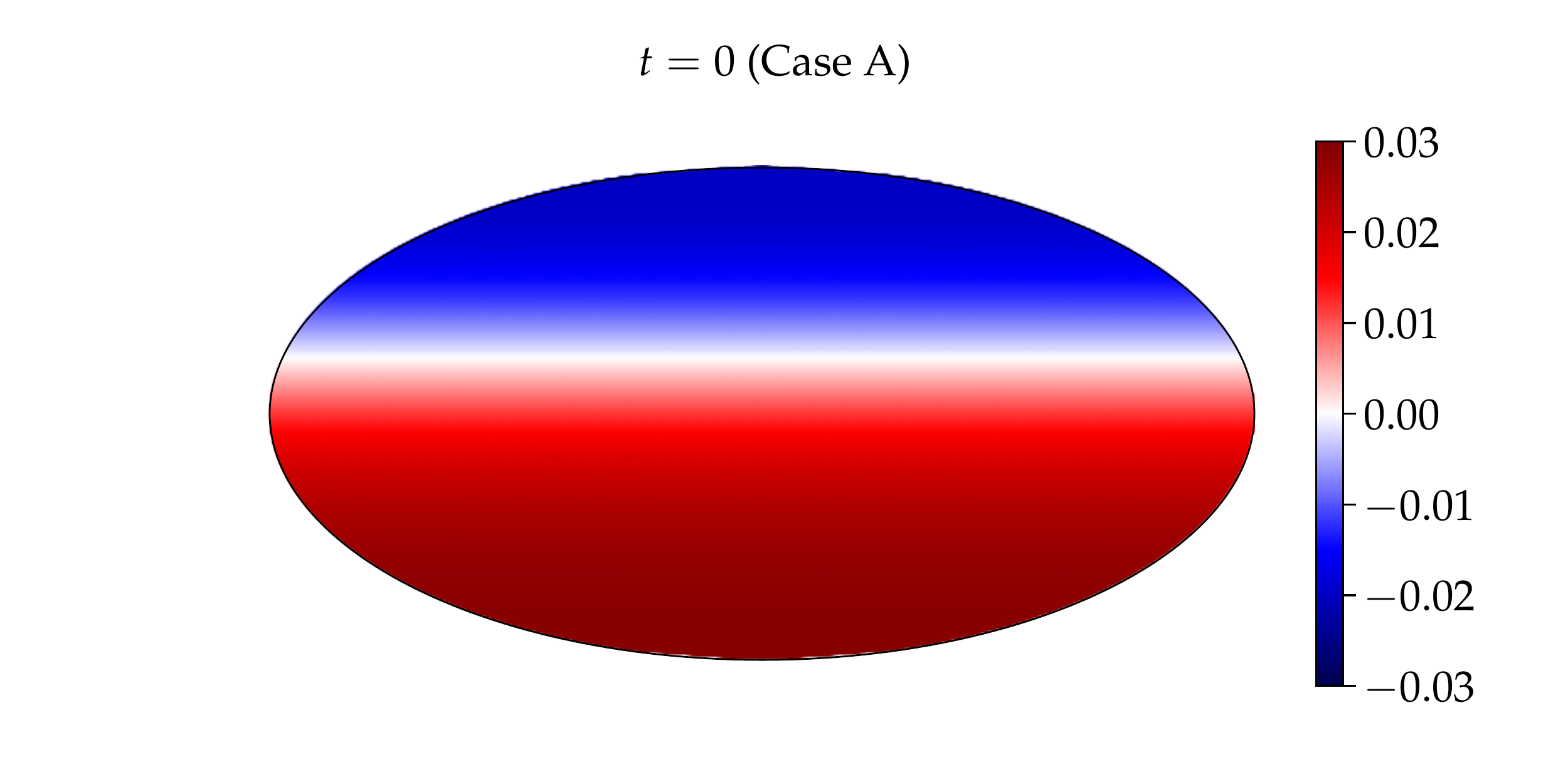}
\includegraphics[width=0.49\textwidth]{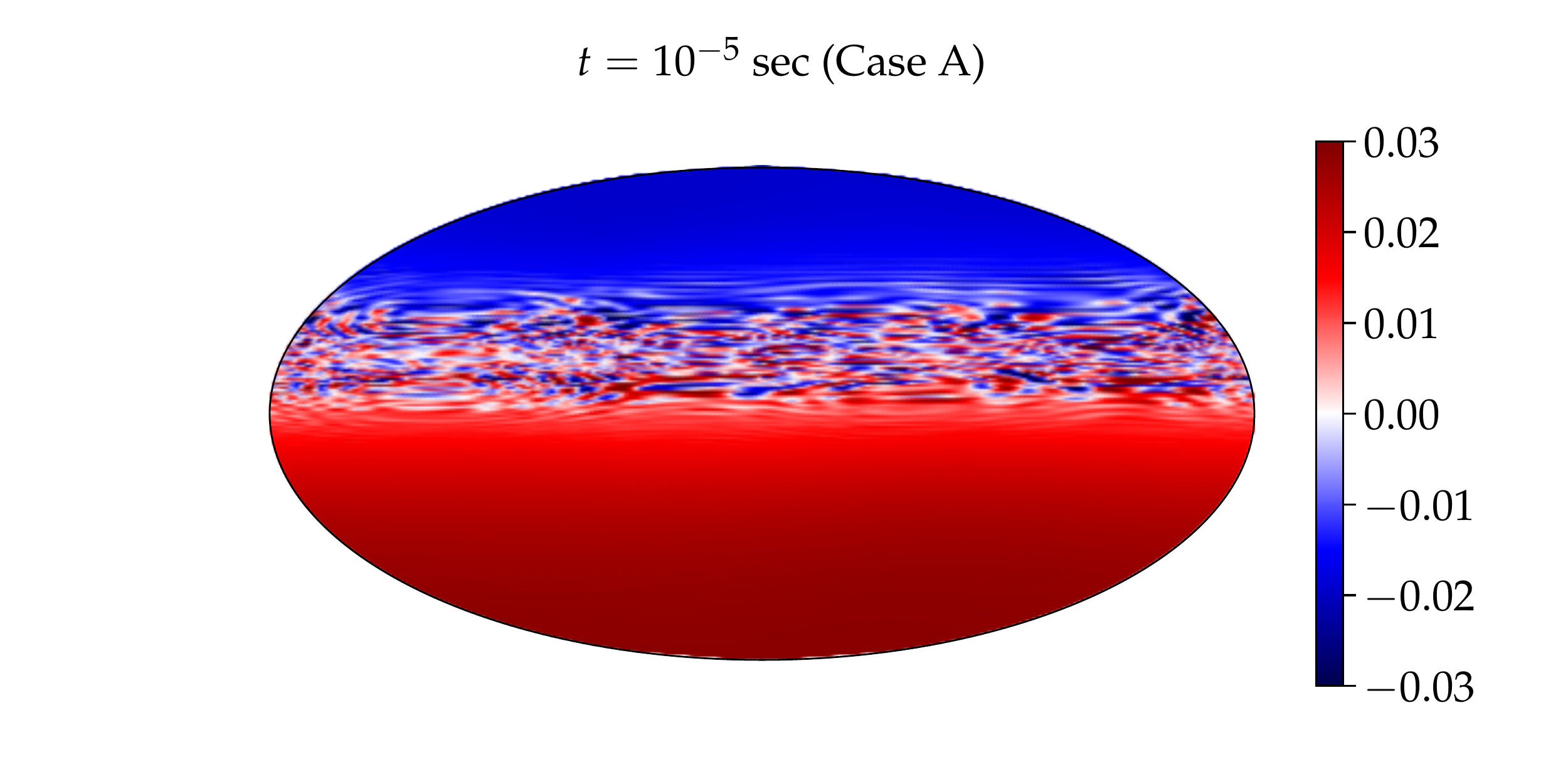}\\
\includegraphics[width=0.49\textwidth]{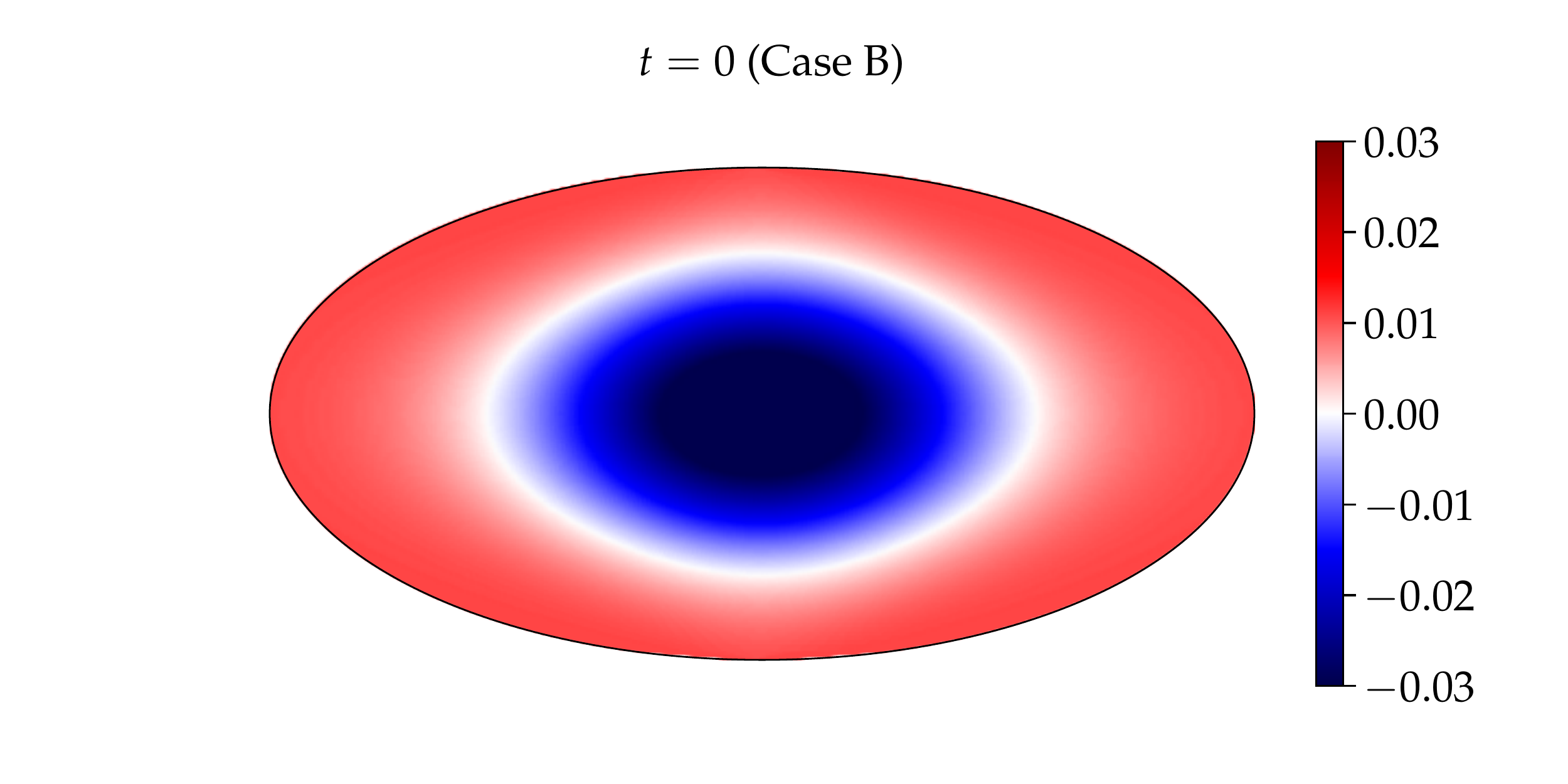}
\includegraphics[width=0.49\textwidth]{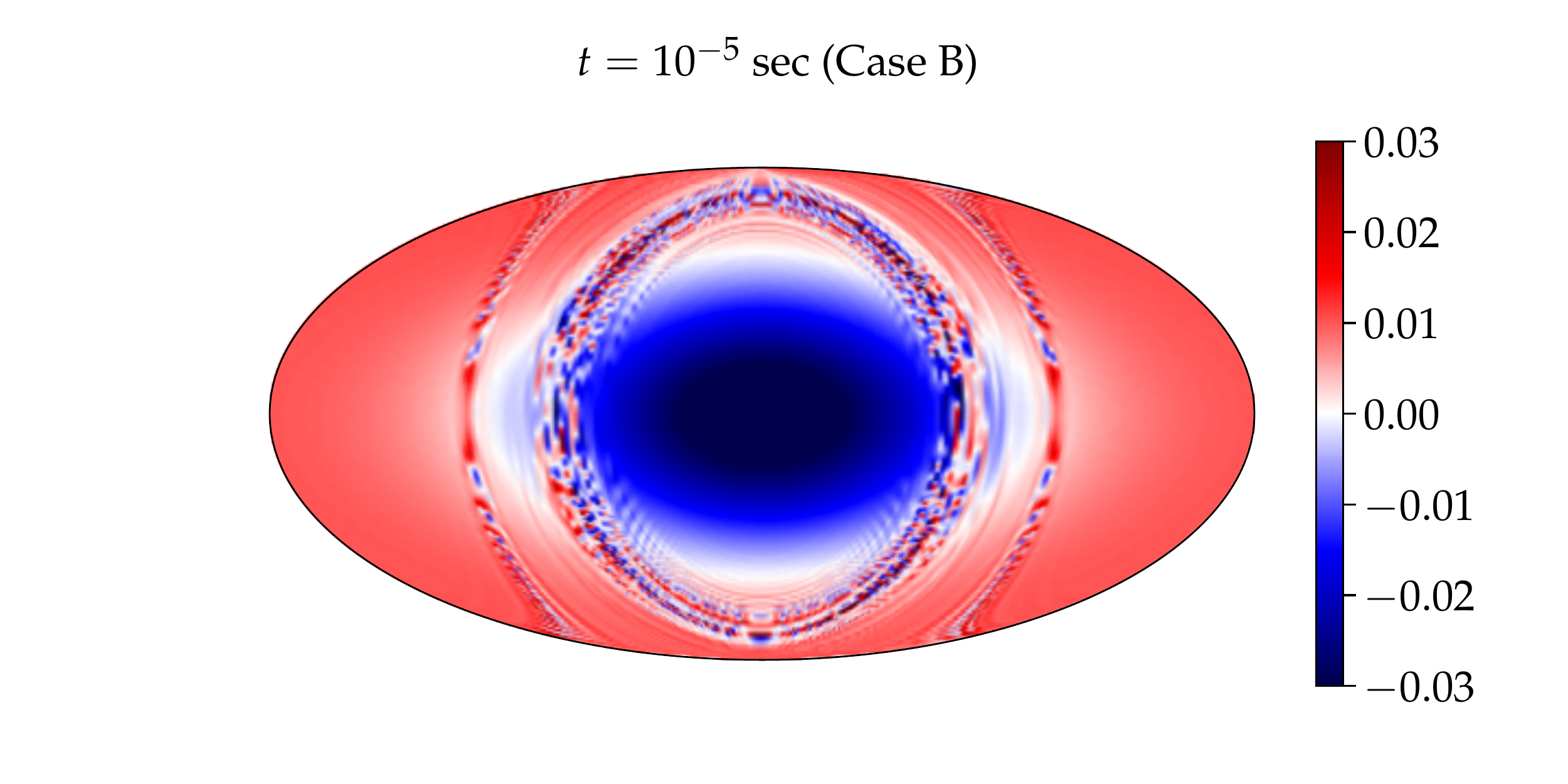}\\
\includegraphics[width=0.49\textwidth]{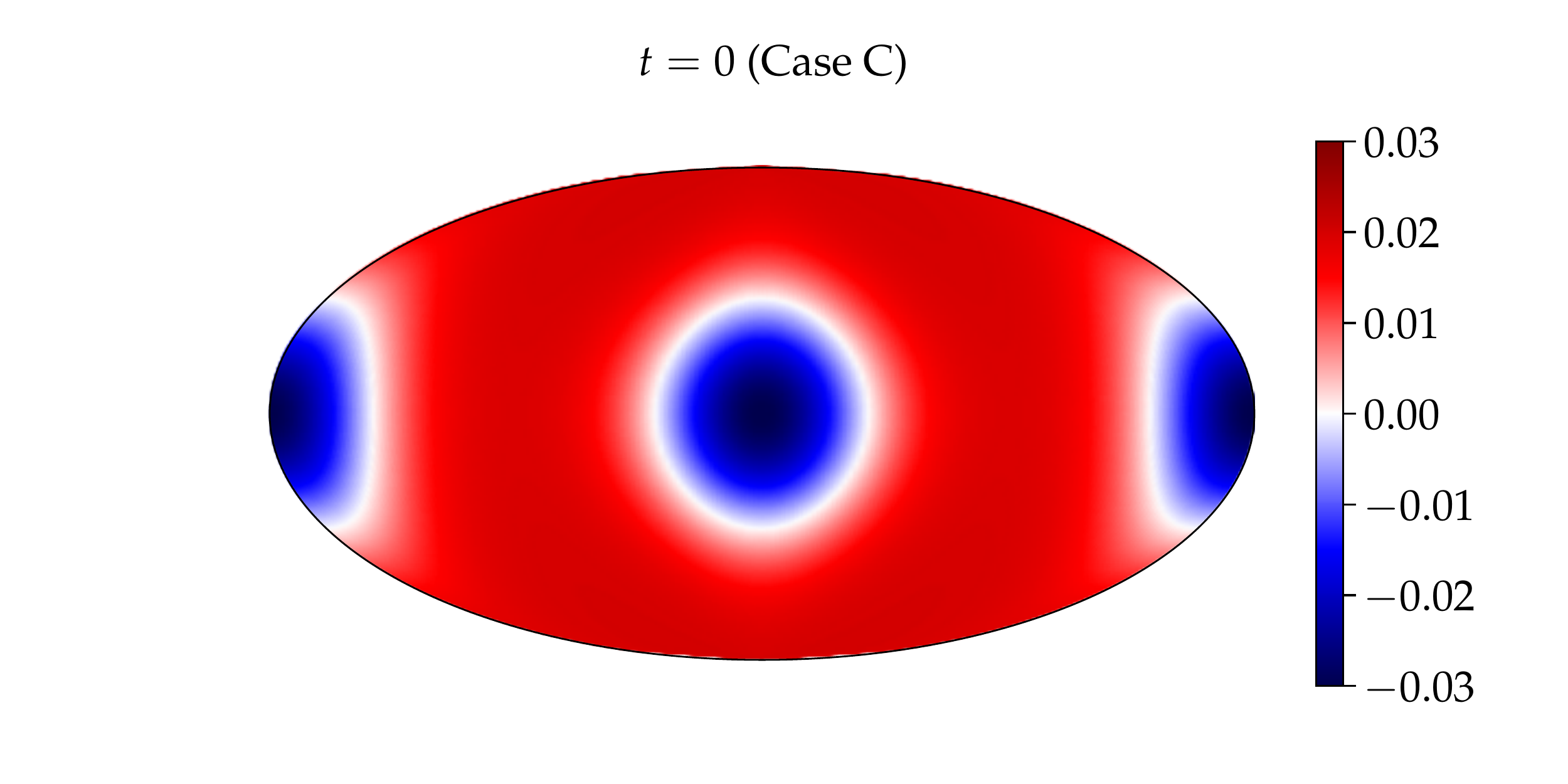}
\includegraphics[width=0.49\textwidth]{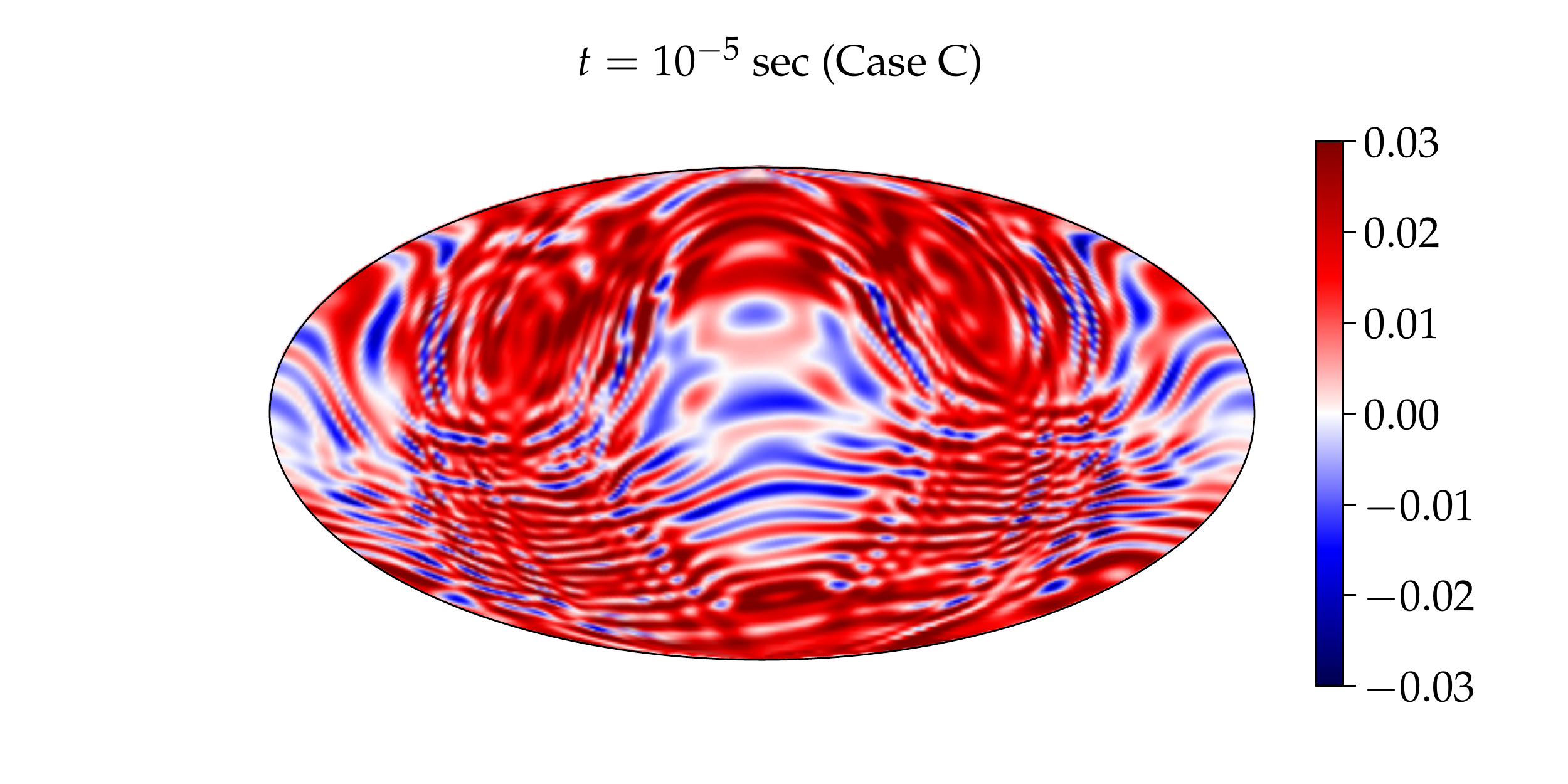}\\
\includegraphics[width=0.49\textwidth]{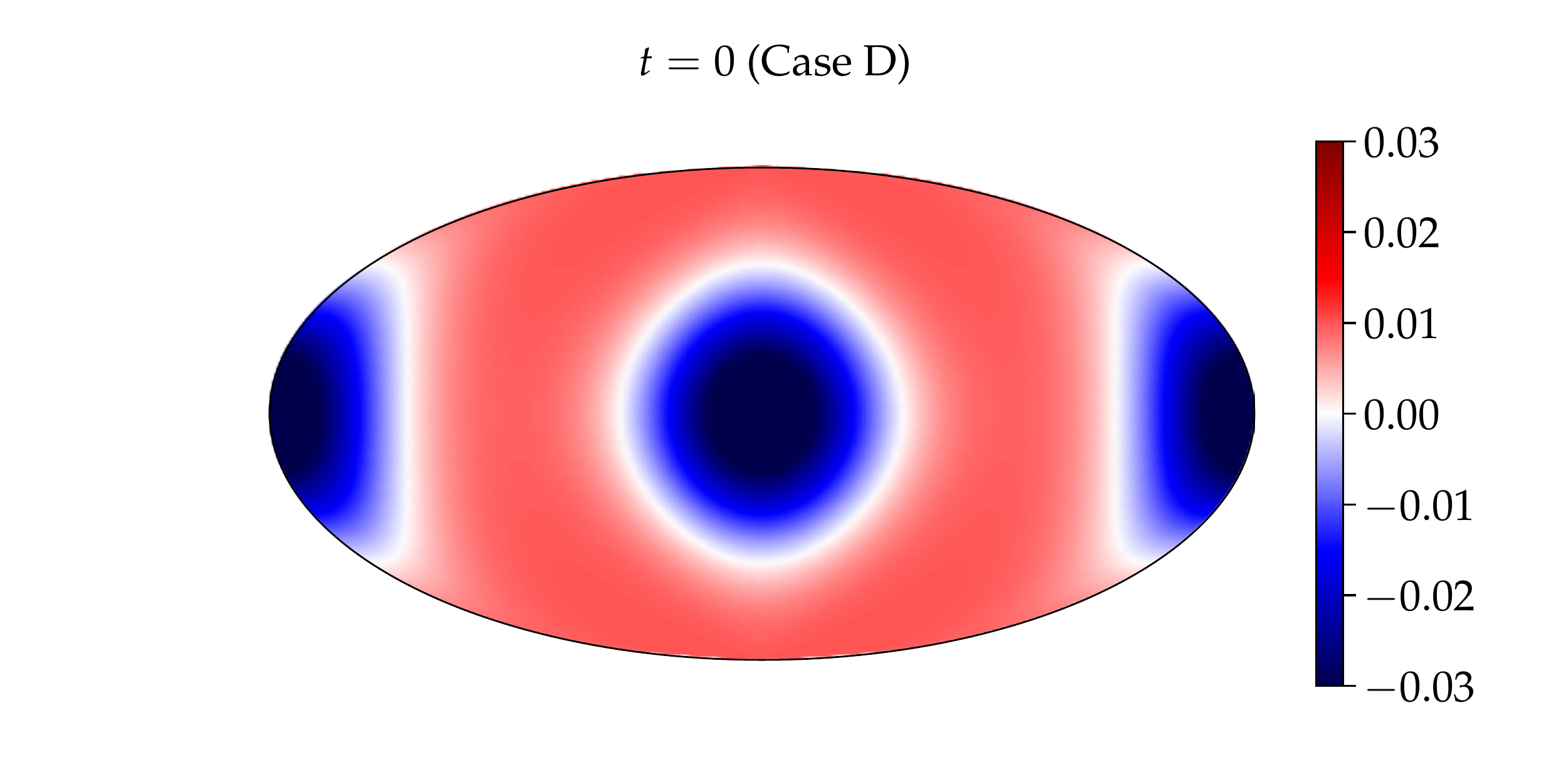}
\includegraphics[width=0.49\textwidth]{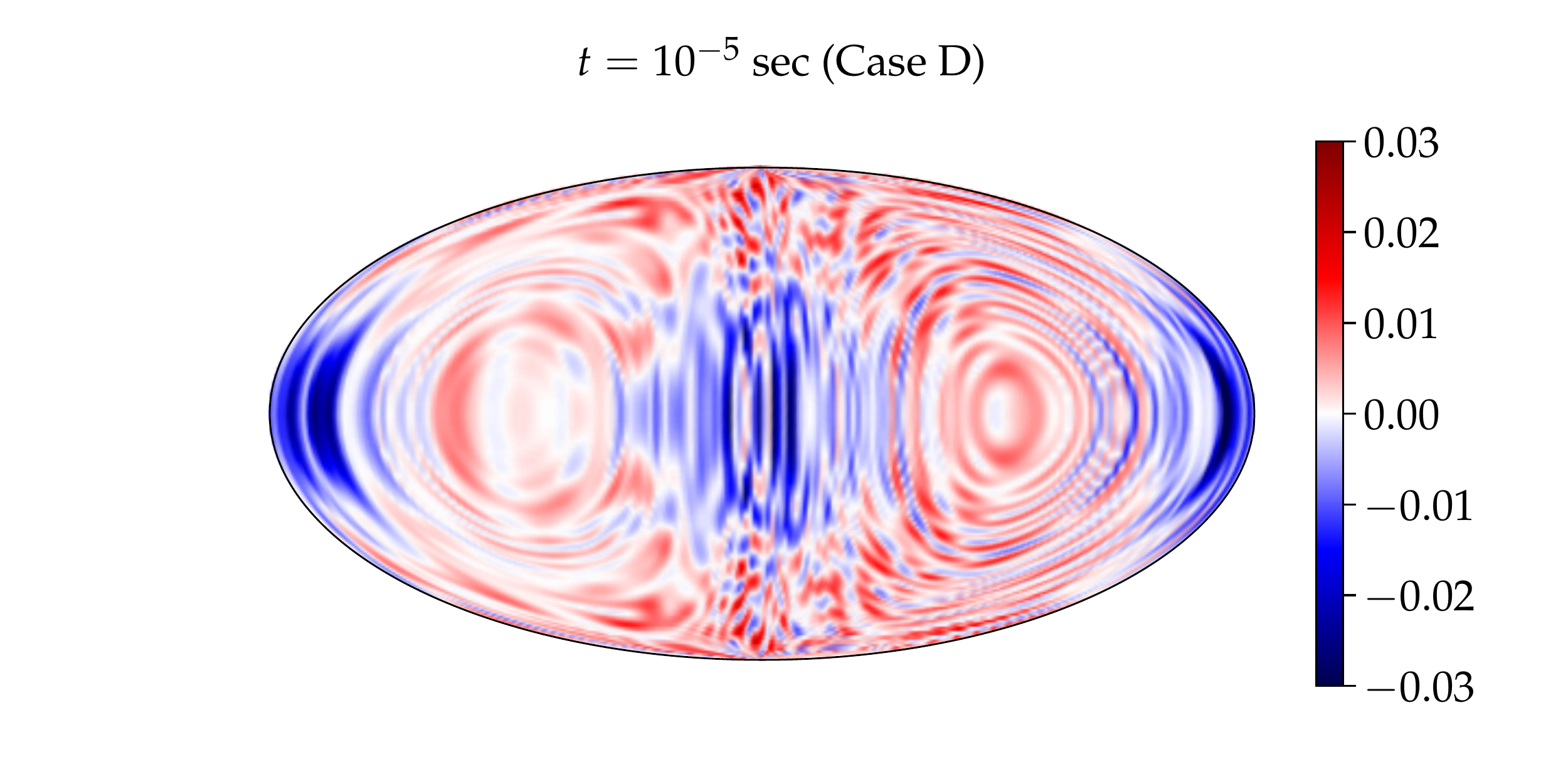}\\
\caption{{\it Left panels:} Mollweide projections of the angular distribution of the initial  $\rho_{ee} - \bar{\rho}_{ee}$   for Cases A--D  from top to bottom, respectively (see Eqs.~\ref{casea}--\ref{casee}). On each skymap, the polar variable $\theta$ varies from $0$ at the South pole to $\pi$ at the North pole; while the azimuthal variable $\phi$ runs from $0$ on the left hand side of the map to $2\pi$ on the right hand side. The white region in each map shows  the  vanishing ELN ring.  {\it Right panels:} Mollweide projections of the angular distribution of the final  $\rho_{ee} - \bar{\rho}_{ee}$   at $t = 10^{-5}$~sec. Spontaneous symmetry breaking occurs for all cases and it is especially evident in Case D;  the onset of flavor mixing does not always follow the ELN rings, as clearly evident in Cases B--D.
}
\label{Fig2}
\end{figure*}

To better appreciate the differences among Cases A--D,  the  angular distribution of ${\rho}_{ee}-\bar{\rho}_{ee}$ is shown for all configurations in the left panels of Fig.~\ref{Fig2}. Cases A and B have one single ELN ring, but  Case A shows one crossing in $\theta$ and it is azimuthally symmetric (i.e.~this configuration can be directly mapped onto a one-dimensional angular distribution similar to the ones  commonly adopted in the literature). 
Despite the different orientation of their respective ELN ring, Cases A and B have been selected for a direct comparison with the existing literature, where it has been shown that the instability type depends on the kind of ELN crossing and the flavor outcome is strictly related to the width and height of the ELN crossing~\cite{Yi:2019hrp,Martin:2019gxb,Tamborra:2020cul}.

Cases C and D have two ELN rings, with Case D having larger rings than Case C (see also Tab.~\ref{Tab1}). 
 In the limit of azimuthal symmetry, it was found that  distributions with two ELN crossings  are not necessarily unstable~\cite{Capozzi:2019lso,Bhattacharyya:2021klg}.
 However, as we  show in the following, the magnitude of the flavor instability in the linear regime is not an indication of the amount of flavor conversion, as also found in other investigations of  fast flavor conversion in the non-linear regime, see e.g.~\cite{Shalgar:2021wlj, Shalgar:2020xns}.
 
It is also worth noticing that Cases A--D involve symmetries in the initial conditions (see left panels of Fig.~\ref{Fig2}); for example, there is a reflection symmetry around $\phi=\pi$ and for all cases, except for Case A, there is a reflection symmetry around $\theta=\pi/2$. However, as shown in the following,  these initial symmetries are broken through flavor mixing.

\section{Equations of motion}
\label{setup}

The evolution of neutrinos and antineutrinos is governed by the Heisenberg equations:
\begin{eqnarray}
i\frac{d\rho}{dt} = [H,\rho]\ \mathrm{and}\ 
i\frac{d\bar{\rho}}{dt} = [\bar{H},\bar{\rho}]\ ;
\end{eqnarray}
the Hamiltonian includes three components corresponding to the vacuum, matter and self-interaction terms:
\begin{eqnarray}
H &=& H_{\textrm{vac}} + H_{\textrm{mat}} + H_{\nu\nu}\ ,\\ 
\bar{H} &=& -H_{\textrm{vac}} + H_{\textrm{mat}} + H_{\nu\nu}\ .
\end{eqnarray}
The three terms of the Hamiltonian are defined as
\begin{eqnarray}
H_{\textrm{vac}} &=& \frac{1}{2 E} U \textrm{diag} (0, \delta m^{2}, \Delta m^{2}) U^{\dagger}\ ,\\
H_{\textrm{mat}} &=& \textrm{diag} (\sqrt{2} G_{\textrm{F}} n_{e}, 0, 0)\ ,\\
H_{\nu\nu} &=& \mu \int (\rho - \bar{\rho}) (1-v\cdot v^{\prime}) d^{3}v^\prime\ . \label{eq:Hnunu}
\end{eqnarray}
Here $U(\theta_{12}, \theta_{13}, \theta_{23})$ is the  $3 \times 3$ Pontecorvo-Maki-Nakagawa-Sakata matrix expressed as a function of the three mixing angles, $\delta m^{2}$ and $\Delta m^{2}$ are the mass-squared differences of neutrinos, $G_{\textrm{F}}$ is the Fermi constant, $n_{e}$ is the effective number density of electrons, $\mu$ is the strength of neutrino self-interaction, $v$ is the velocity of the neutrinos under consideration and $v^{\prime}$ represents the velocity of the neutrinos in the medium.

The $(1-v\cdot v^{\prime})$ term in $H_{\nu\nu}$ can be re-written  in terms of $\theta$ and $\phi$ explicitly using: 
\begin{eqnarray}
v\cdot v^{\prime} &=& \cos\theta \cos\theta^{\prime} + \sin\theta \sin\theta^{\prime} (\cos\phi \cos\phi^{\prime} + \sin\phi \sin\phi^{\prime})\ . \nonumber \label{eq:Hangles}
\end{eqnarray}
It is worth noticing that since the inner product  in $H_{\nu\nu}$ is not symmetric in $\theta$ and $\phi$, the  flavor evolution cannot be decomposed into polar and azimuthal modes; nevertheless, it is possible to rotate the location of the ELN rings across the  surface spanned by the polar and azimuthal angles. Moreover, the term $v\cdot v^{\prime}$ in the Hamiltonian highlights the occurrence of symmetries in $H_{\nu\nu}$ that are not necessarily the same as the ones of the ELN rings.
By taking into account Eq.~\ref{eq:Hangles},  $H_{\nu\nu}$ can be split in  four terms,  that are helpful in identifying the symmetries of the system as further discussed in the following:
\begin{eqnarray}
H_{\nu\nu}=h_{0}+h_{1}\sin\theta\cos\phi+h_{2}\sin\theta\sin\phi+h_{3} \cos\theta; 
\nonumber\\
\label{eq14}
\end{eqnarray}
e.g., if  $h_{3}$ is initially zero, this is an indication of a discrete reflection symmetry of the Hamiltonian across $\theta=\pi/2$; while the dynamical generation of $h_{3}$  can only be a result of spontaneous breaking of this discrete symmetry.

In the numerical simulations, we assume $\delta m^{2} = 7.53 \times 10^{-5}$~eV$^{2}$, $\Delta m^{2} = 2.453 \times 10^{-3}$~eV$^{2}$~\cite{Zyla:2020zbs},  $\mu = 2.5 \times 10^{4}$ km$^{-1}$,  and only consider  one energy mode $E= 50$~MeV. We ignore the matter term and compensate for that by using $\theta_{12} = \theta_{13} = \theta_{23}= 10^{-3}$~rad (this approximation  has been shown to hold in the three flavor case~\cite{Shalgar:2021wlj}). 

We perform the numerical simulations by discretizing the angular distribution over  a grid of $200 \times 200$ bins,  which is uniform over $\cos\theta$ and $\phi$; note that, while this grid choice  samples the polar region more densely,   the equatorial region is not  sparsely sampled.
For   Case A, we have also performed the computation using $500 \times 500$ bins to ensure that numerical convergence is achieved (results not shown here). 
In the numerical solution of the neutrino equations of motion, the seed to trigger the symmetry breaking modes is not put by hand, but  it is provided by the inaccuracy in the computation of the  trigonometric functions, which do not exactly satisfy the condition $\sin(\phi)+\sin(-\phi)=0$. For completeness, the evolution of the angle-averaged $\nu_e$ flavor content is reported in Appendix~\ref{sec:symmbreak}.

\section{Flavor evolution in the non-linear regime}
\label{numerical}
\begin{figure}
\includegraphics[width=0.49\textwidth]{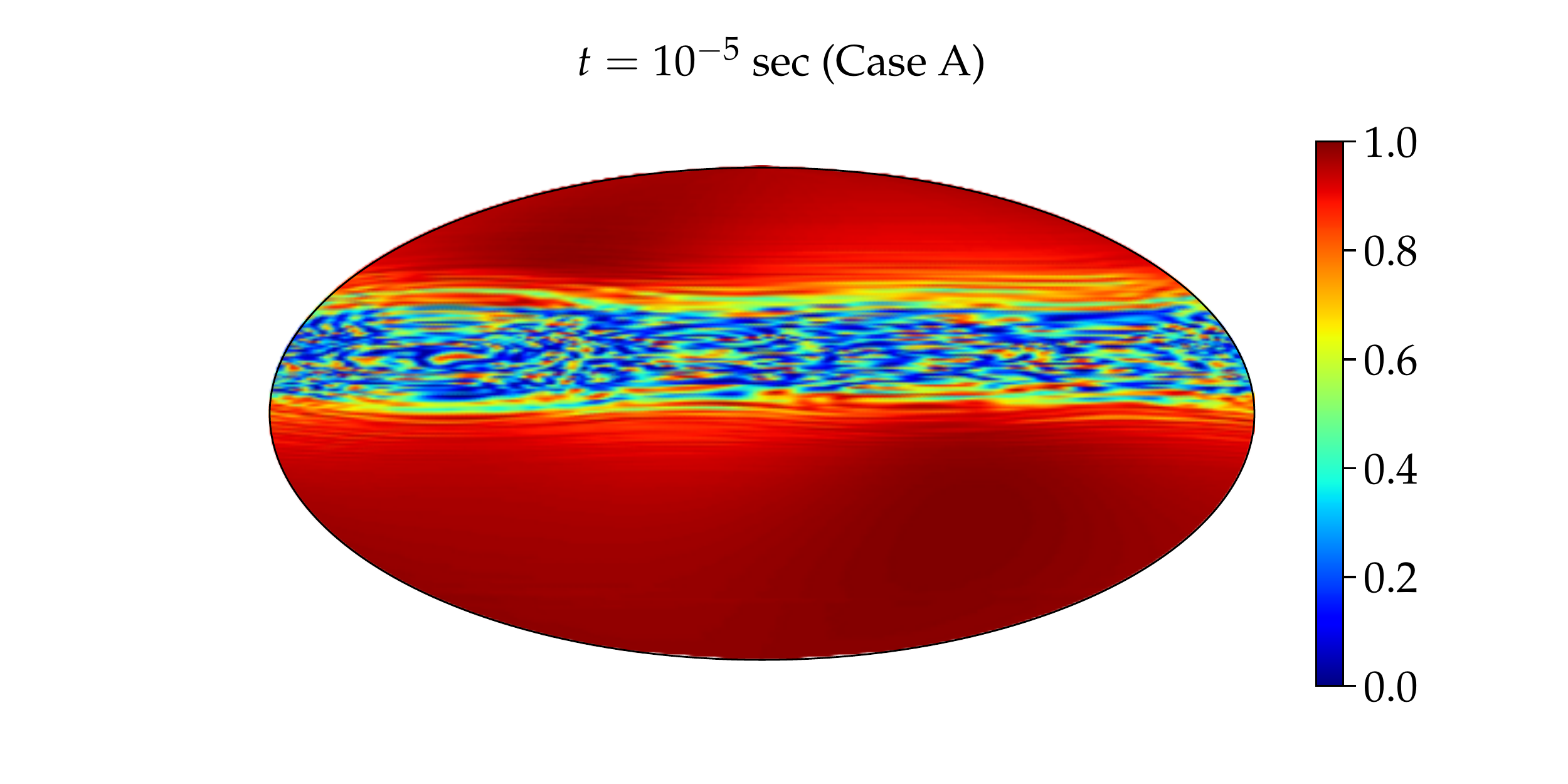}
\includegraphics[width=0.49\textwidth]{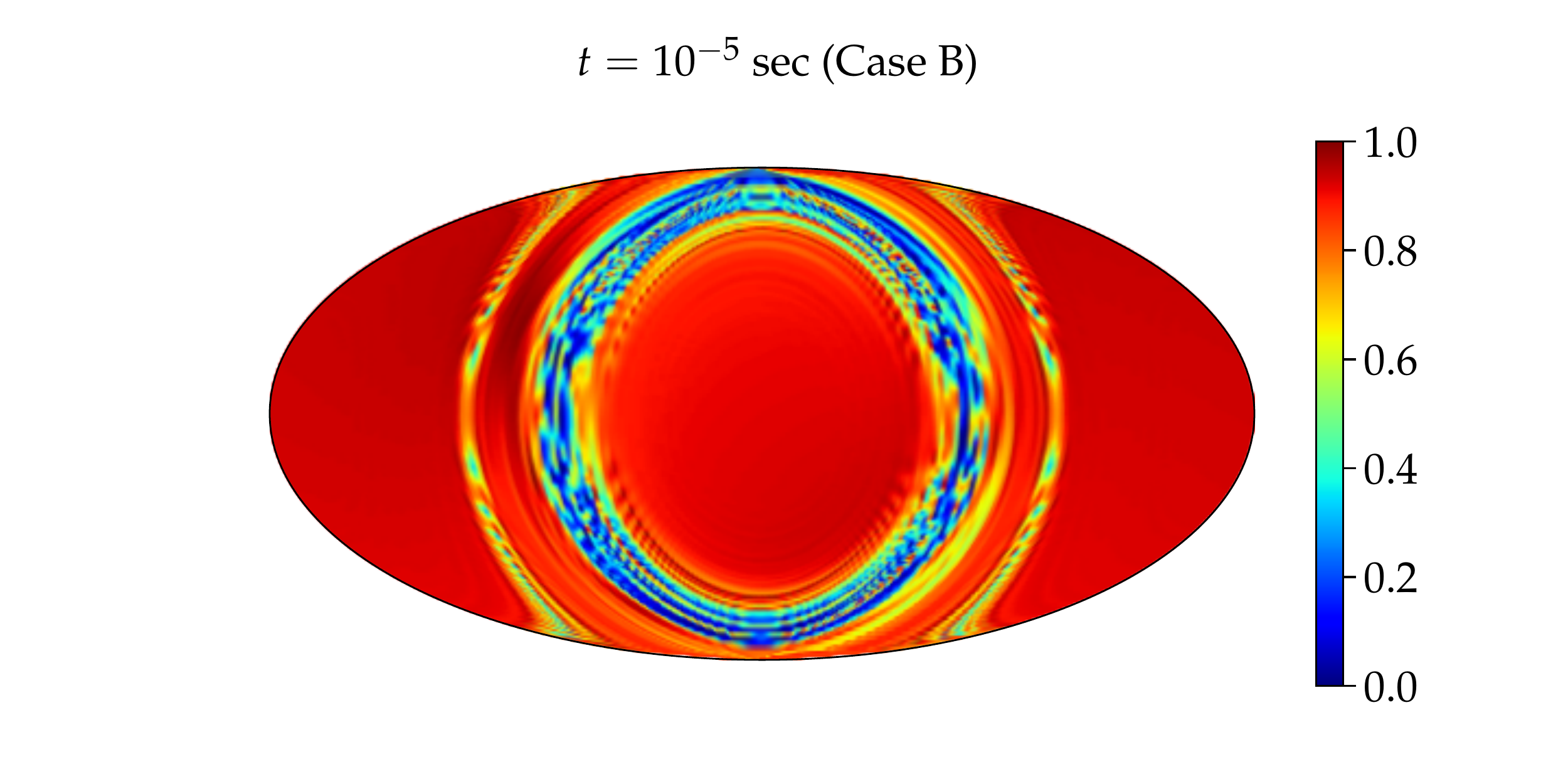}
\includegraphics[width=0.49\textwidth]{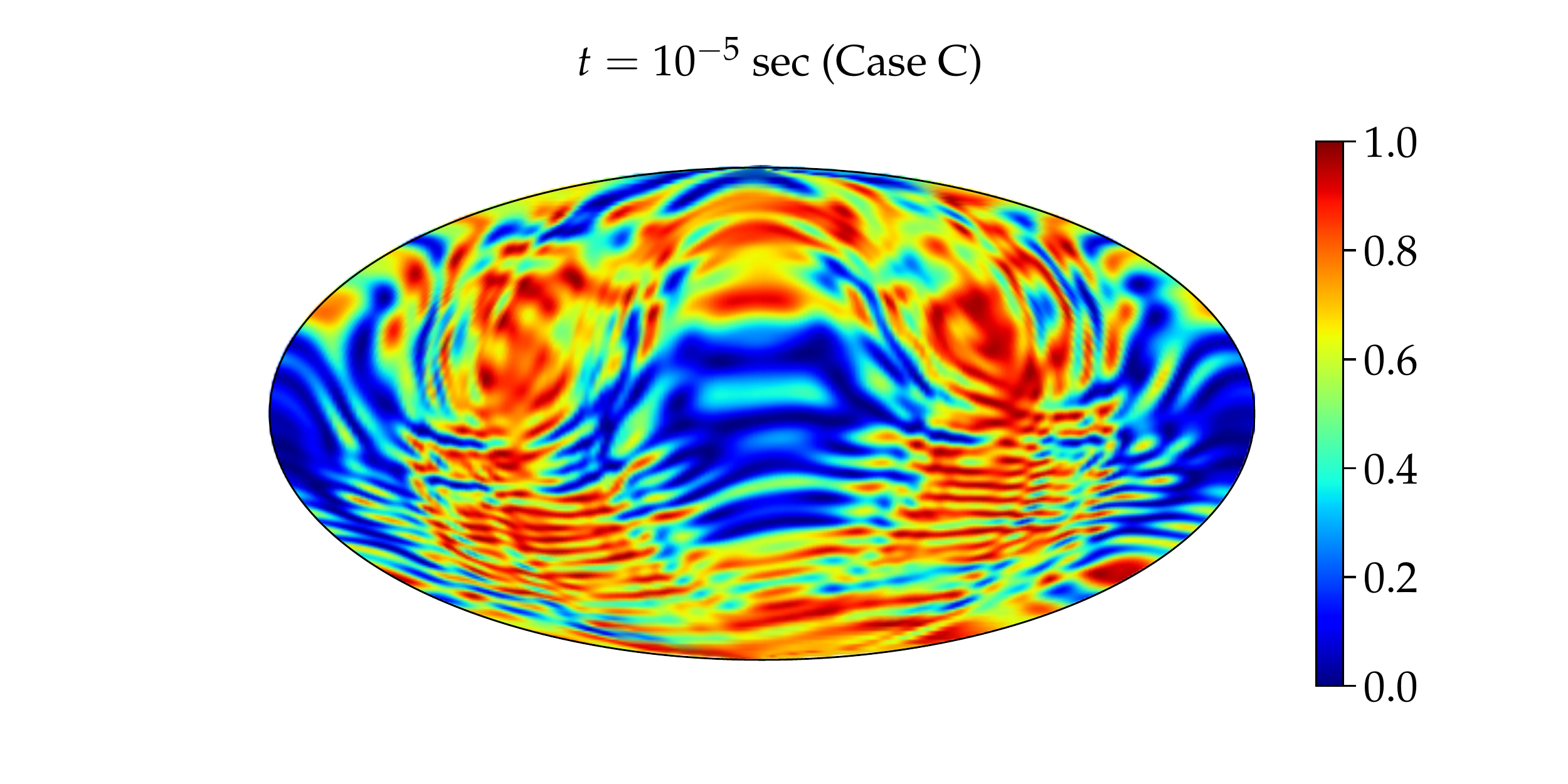}
\includegraphics[width=0.49\textwidth]{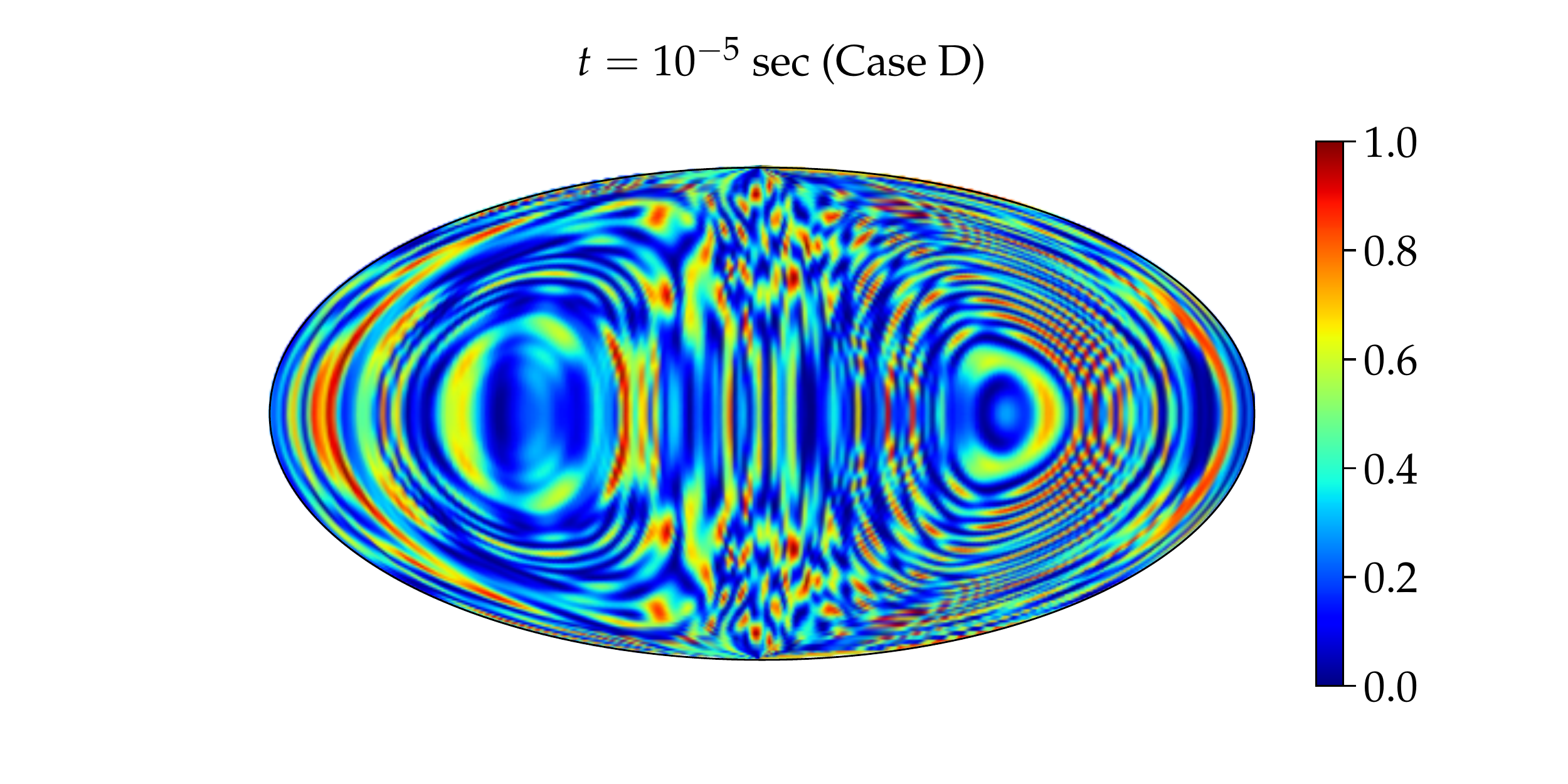}
\caption{Mollweide projections of the ratio between the initial and final content of $\nu_e$  (defined as in Eq.~\ref{eq:Pee}) at $t = 10^{-5}$~sec for Cases A--D from top to bottom, respectively. Maximal flavor mixing is achieved for all Cases for certain regions across the Mollweide map. 
}
\label{Fig3}
\end{figure}

The top right panel of Fig.~\ref{Fig2} shows the   map of  $\rho_{ee} - \bar{\rho}_{ee}$ of Case A at  $t=10^{-5}$~sec. Although we start with a system that has azimuthal symmetry,  the initial reflection symmetry around $\phi=\pi$  is broken spontaneously (see also the \href{https://sid.erda.dk/share_redirect/hwp0WKuuk7/index.html}{animations} in the Supplemental Material).  The symmetry breaking is particularly visible in the top panel of Fig.~\ref{Fig3}, showing  the  Mollweide projection of $P_{ee}$, where the latter is defined as 
 the ratio between the final and initial  content of $\nu_e$: 
\begin{eqnarray}
\label{eq:Pee}
P_{ee}  = \frac{\rho_{ee}(\cos\theta,\phi,t)}{\rho_{ee}(\cos\theta, \phi, t=0)}\ ;
\end{eqnarray}
  there are regions across the Mollweide map where  near maximal flavor transformation is achieved (see  Appendix~\ref{sec:symmbreak} for additional details). The right top panel of Fig.~\ref{Fig2} and the top panel of Fig.~\ref{Fig3} suggest  that the onset of the flavor evolution for Case A starts in the proximity of the  ELN ring, which happens to be circular in shape. This finding is in agreement with work on 
  azimuthally symmetric cases~\cite{Yi:2019hrp,Martin:2019gxb,Tamborra:2020cul}. Flavor mixing then spreads in the upper hemisphere in the non-linear regime. 

Case B is another configuration with one ELN ring (second  panel on the left of Fig.~\ref{Fig2}). As shown in the   corresponding panel on the right of Fig.~\ref{Fig2} and in the second   panel of Fig.~\ref{Fig3}  (as well as in the \href{https://sid.erda.dk/share_redirect/hwp0WKuuk7/index.html}{animations} in the Supplemental Material), the onset of flavor evolution occurs along a region that is circular in shape, which only intersects the initial  ELN ring, while not being  aligned with it. This result has been overlooked  in azimuthally symmetric configurations (see, e.g., Case A for comparison). 
The reason for such a trend  can be easily understood by looking at the Hamiltonian of the system  (Eqs.~\ref{eq:Hnunu} and \ref{eq:Hangles}); the  Hamiltonian is only related to the ELN angular distribution through the  integration over the polar and azimuthal angles, hence $H_{\nu\nu}$ can have symmetries that  the initial ELN distribution does not have;    in Case B,  $h_1 \neq 0$ and $h_1  \gg h_{0,2,3} \simeq 0$ at $t = 0$; then $h_1$ reaches the non-linear regime  followed by $h_0$ and dominates the onset of the non-linear phase (see Appendix~\ref{sec:symmbreak} for more details). As a consequence, flavor mixing develops along a circle, but does not  track the ELN ring.

The phenomenology of fast flavor conversion becomes even more interesting when two ELN rings are considered (Cases C and D). In fact, even if the two initial configurations appear to be very similar except for the overall opposite ELN sign (see  Table~\ref{Tab1} and  the two panels on the bottom left of Fig.~\ref{Fig2}), the final $\rho_{ee} - \bar{\rho}_{ee}$  configurations  are very different in the  two panels on the bottom right of Fig.~\ref{Fig2}.
For Case C, the onset of flavor mixing occurs along one ring crossing the North and the South poles of the map of angular distribution (see \href{https://sid.erda.dk/share_redirect/hwp0WKuuk7/index.html}{animations}  in the Supplemental Material), then  the flavor evolution is such that the final flavor configuration  mixing happens to resemble the shape of the initial  ELN rings. 

In Case D, the ELN pattern after flavor mixing is completely different with respect to the initial ELN configuration and the angular regions with significant flavor evolution are not correlated in an obvious manner with the regions where  the ELN rings are initially located, as also visible from the \href{https://sid.erda.dk/share_redirect/hwp0WKuuk7/index.html}{animations} in the Supplemental Material. In addition, it is clear from the bottom panel of Fig.~\ref{Fig3}  that full flavor conversion takes place across the whole angular range in Case D.
The difference in the outcome of the flavor evolution is also seen in the evolution of the angle averaged survival probability in Appendix~\ref{sec:symmbreak}. 
The strong symmetry breaking effects observable in Case D are not determined by the small magnitude of the ELN per se---the latter has indeed  been employed in various sets of initial configurations, see e.g.~Refs.~\cite{Padilla-Gay:2020uxa,Wu:2017qpc};  it is also not determined by  the negative sign of the total ELN, in fact we have explored other non-azimuthally symmetric configurations with overall  negative ELN without finding dramatic spontaneous symmetry breaking effects (results not shown here).

It is important to stress that such symmetry breaking is different in nature with respect to the one observed in the context of slow neutrino self-interaction~\cite{Raffelt:2013rqa, Duan:2014gfa, Abbar:2015mca}, where the of azimuthal symmetry breaking has been investigated in connection to  breaking of spatial symmetry.
Fast flavor evolution can occur over small length scales and the region over which flavor mixing occurs can be approximated to have homogeneous initial conditions. Slow neutrino self-interactions, on the other hand, occur over larger length scales and have been investigated mostly in the context of neutrino-bulb model; as such,   the initial conditions are not homogeneous resulting in a system that cannot be solved in a self-consistent manner if we only break the azimuthal symmetry while preserving the initial spatial symmetry.  
In the case of fast flavor mixing, the angular symmetry breaking is purely driven by the angular distributions of neutrinos.  The symmetry breaking becomes fully evident in the non-linear regime,  emphasizing the need for going beyond the linear stability analysis in the context of fast flavor mixing in agreement with Refs.~\cite{Johns:2019izj, Shalgar:2021wlj, Shalgar:2020wcx, Shalgar:2020xns, Chakraborty:2019wxe, Capozzi:2020kge}.

\section{Conclusions}
\label{conclusions}
In this paper, for the first time, we explore the flavor conversion physics in three flavors and by including the polar and azimuthal angular distributions.  Under the assumption of azimuthal symmetry, it is usually believed that  fast flavor conversion develops in the angular bins in the proximity of the  electron lepton number (ELN) ring.

When the  assumption of azimuthal symmetry is  relaxed, we  demonstrate that  the onset of flavor mixing does not follow the ELN rings; flavor mixing   develops along angular regions that are circular in shape and that intersect the ELN ring, but do not always fully overlap with the ELN ring. This is an interesting feature of fast conversion. The correlation between the location of the ELN ring and the one of the onset of flavor evolution   in the azimuthally symmetric configurations seems to be a  peculiar feature of these configurations.
 This phenomenon is due to the fact that  the $\nu$--$\nu$ Hamiltonian  can be azimuthally symmetric, even if the  ELN distribution is not.

As  flavor evolution enters the non-linear regime, the angular regions affected by mixing are  not restricted to the initial circular region; all the symmetries present in the initial ELN configuration are broken. In the examples investigated in this work, even the 
 discrete reflection  symmetry imposed in the initial  configuration of our system is spontaneously broken by the floating-point error in the numerical simulation. 

Intriguingly, we find  configurations for which the angular distribution after flavor conversion  is entirely different from the initial one and flavor mixing   affects all angular modes. This effect does not manifest in the context of slow self-interactions and it is observed to occur for fast flavor mixing for the first time. These findings reinforce the need for  more substantial work on the fast flavor evolution in the non-linear regime.

We  focused on the flavor evolution in the case of a spatially homogeneous neutrino gas; however, as the symmetry breaking in the neutrino flavor conversion starts manifesting in the non-linear regime, the assumption of homogeneity  may eventually  play an important role. In addition,  we have ignored the effect of direction changing collisions on the system, which could further aid the symmetry breaking effects. Our work highlights that it is necessary to carry out a thorough investigation of the physics linked to spontaneous symmetry breaking due to flavor mixing within realistic astrophysical systems and in the non-linear regime.

\acknowledgments
We are grateful to the Villum Foundation (Project No.~13164), the Danmarks Frie Forskningsfonds (Project No.~8049-00038B),  the MERAC Foundation, and the Deutsche Forschungsgemeinschaft through Sonderforschungbereich
SFB~1258 ``Neutrinos and Dark Matter in Astro- and
Particle Physics'' (NDM).



\appendix


\setcounter{equation}{0}
\setcounter{figure}{0}
\setcounter{table}{0}
\makeatletter
\renewcommand{\theequation}{A\arabic{equation}}
\renewcommand{\thefigure}{A\arabic{figure}}





\section{Highlights on the development of symmetry breaking} \label{sec:symmbreak}

\begin{figure}[h]
\includegraphics[width=0.49\textwidth]{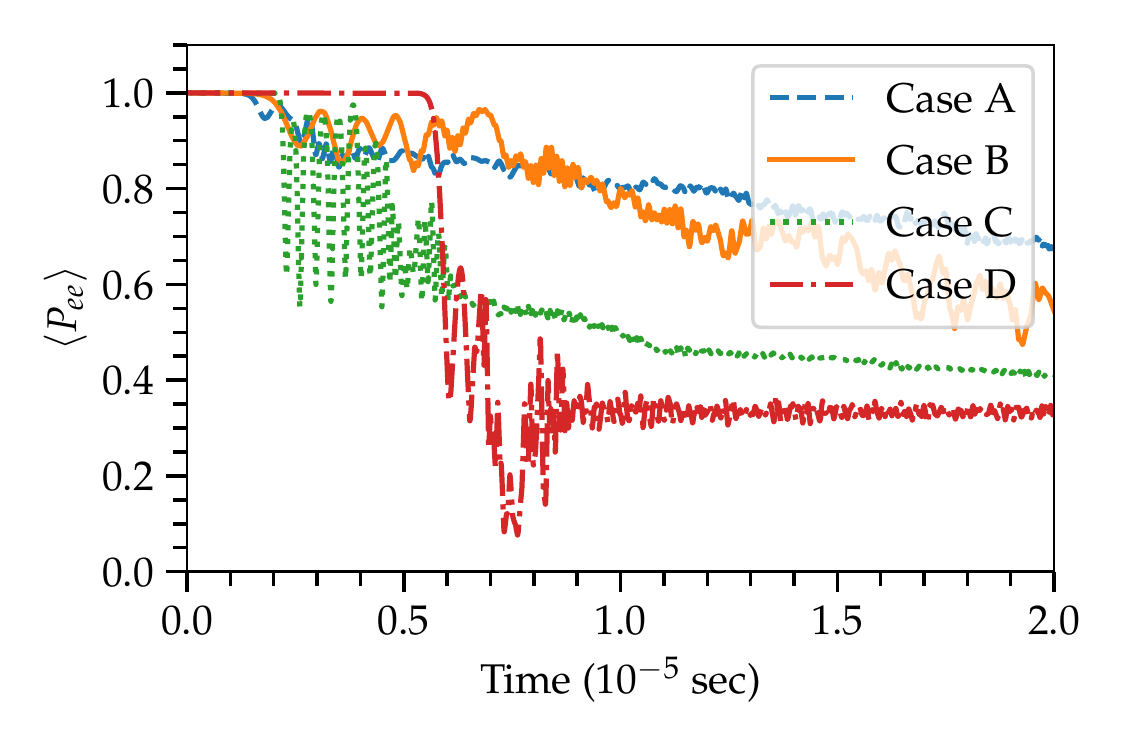}
\caption{Angle averaged ratio between the initial and final content of $\nu_e$    as a function of time for Cases A, B, C, and D. For certain initial configurations, large flavor mixing is achieved.}
\label{Fig4}
\end{figure}

In order to quantify the overall amount of flavor mixing for Cases A--D, Fig.~\ref{Fig4} shows the temporal evolution of the angle averaged ratio between the initial and final content of $\nu_e$: $\langle P_{ee} \rangle =[\int\rho_{ee}(\cos\theta,\phi,t) d\cos\theta d\phi]/[\int\rho_{ee}(\cos\theta,\phi,t=0) d\cos\theta d\phi]$.
 A qualitative difference in the flavor evolution of Cases A and B, as well as C and D, can be seen in Fig.~\ref{Fig4} despite  the very similar angular distributions. In particular, large flavor mixing can be achieved for Cases C and D. 

\begin{figure}[]
\includegraphics[width=0.49\textwidth]{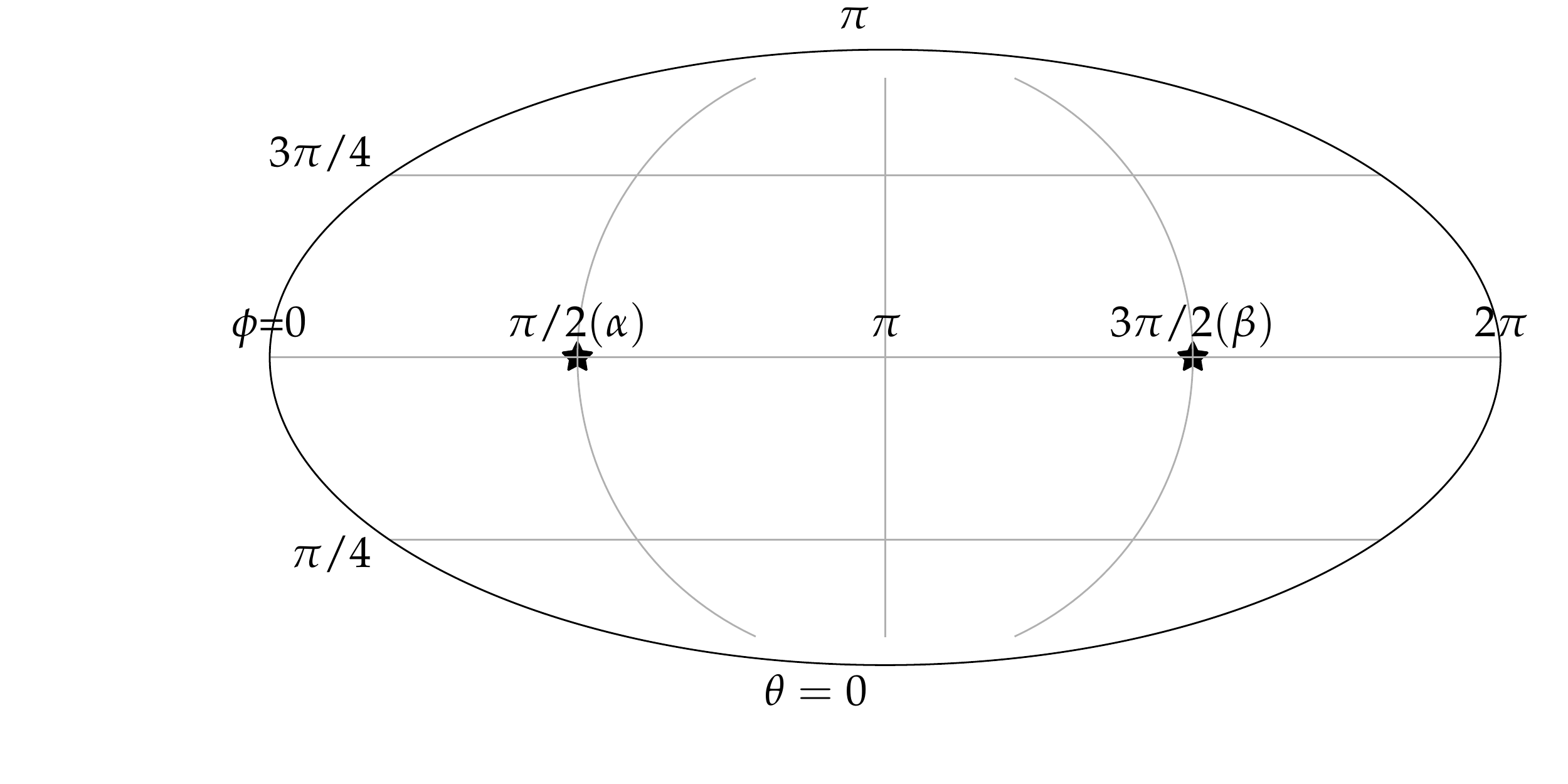}
\caption{Mollweide map illustrating the polar angle ($\theta$) and the azimuthal one ($\phi$) adopted to characterize the angular distributions of (anti)neutrinos. The benchmark points $\alpha$ and $\beta$,  marked by a star, are adopted as representative cases  to investigate the breaking of reflection symmetry around  $(\theta=\pi/2,\phi=\pi)$. 
}
\label{Fig1}
\end{figure}

In this appendix, we intend to provide details on the impact of the symmetry breaking on the flavor evolution. To this purpose, we focus  on  two angular locations, $\alpha \equiv (\theta=\pi/2, \phi=\pi/2)$ and $\beta \equiv (\theta=\pi/2, \phi=3\pi/2)$, displayed  in Fig.~\ref{Fig1} that initially have the same density matrix due to the reflection symmetry with respect to $\phi=\pi$. 

\begin{figure}
\includegraphics[width=0.49\textwidth]{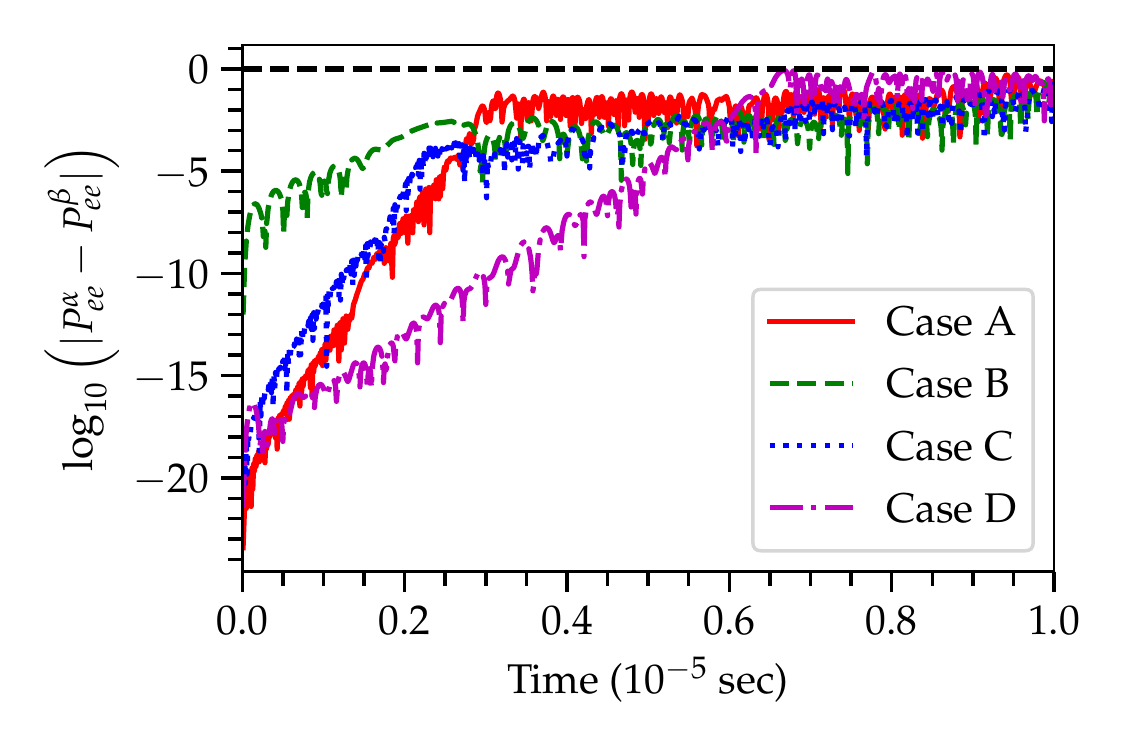}
\caption{Temporal evolution of the  difference in the  ratio between the initial and final content of $\nu_e$ for the points  $\alpha$ and $\beta$ marked  in Fig.~\ref{Fig1}  for Cases A--D.  In all cases, the initial reflection symmetry around $(\theta=\pi/2,\phi=\pi)$  is broken through  mixing. 
}
\label{Fig5}
\end{figure}

The deviation of the $\nu_e$ survival probabilities between $\alpha$ and $\beta$  is a clear indicator  of the symmetry breaking and is displayed in Fig.~\ref{Fig5}.
A close look  at all cases shown in Fig.~\ref{Fig5} highlights that the  linear regime does not give an indication of the magnitude of symmetry breaking that can be expected in the non-linear regime and the time at which such symmetry breaking would  manifest.   

\begin{figure*}[!htb]
\includegraphics[width=0.49\textwidth]{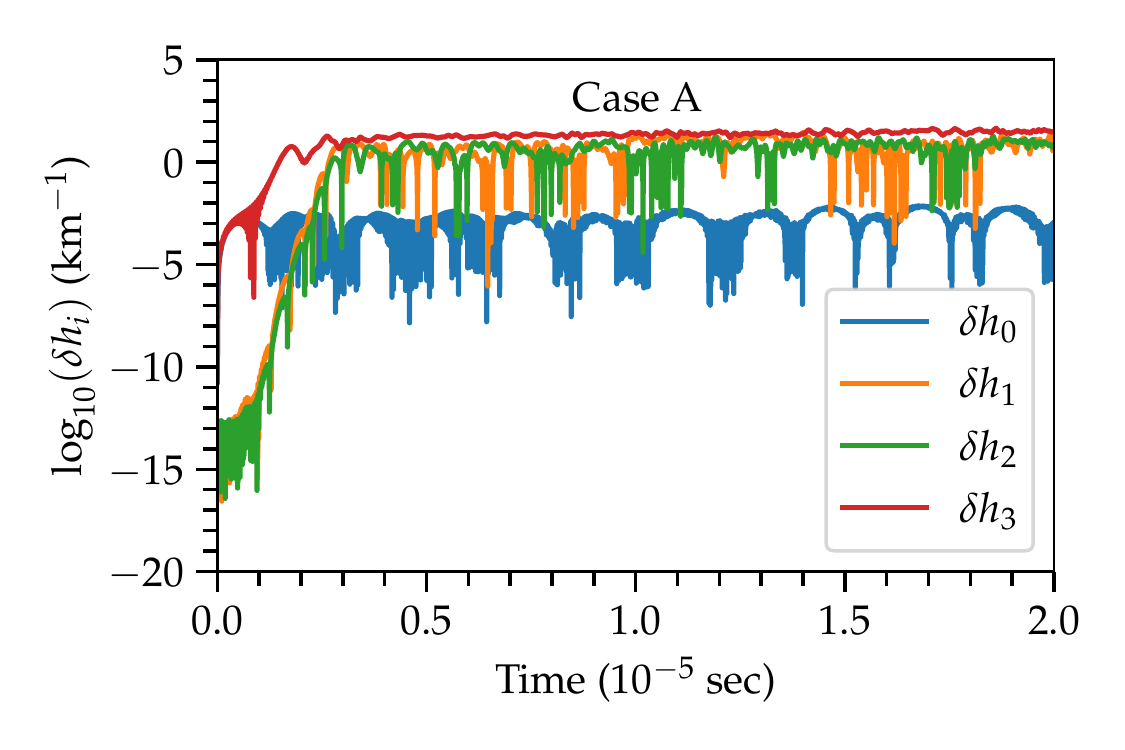}
\includegraphics[width=0.49\textwidth]{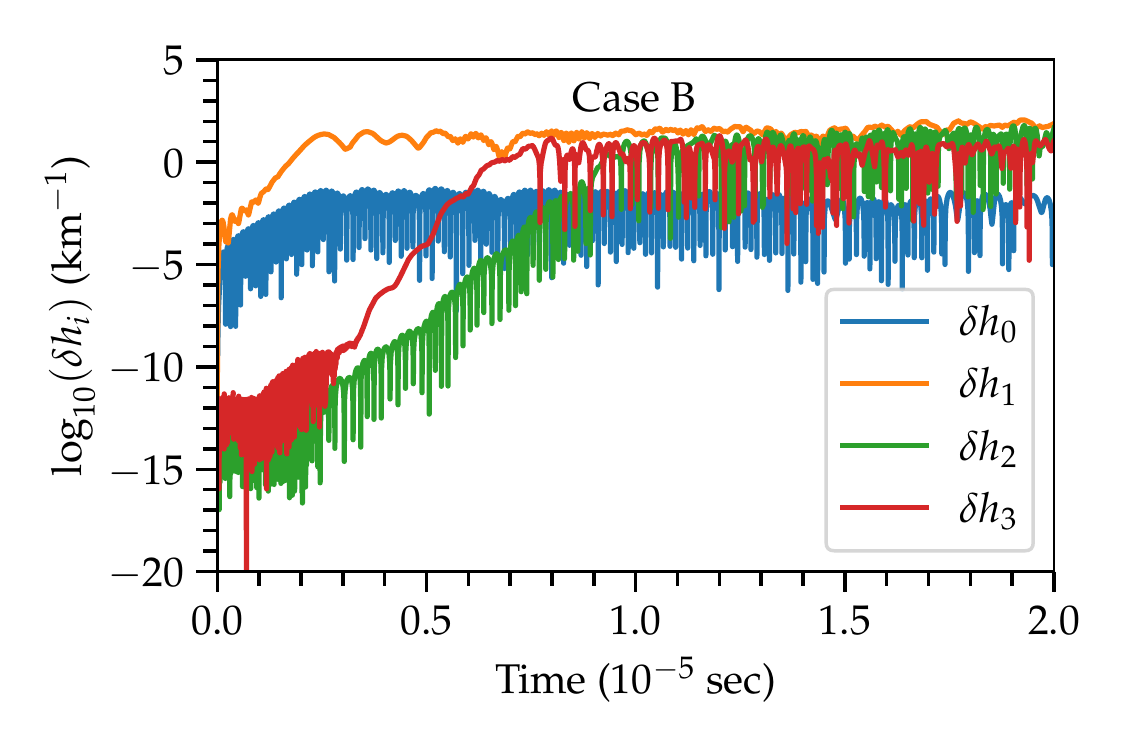}\\
\includegraphics[width=0.49\textwidth]{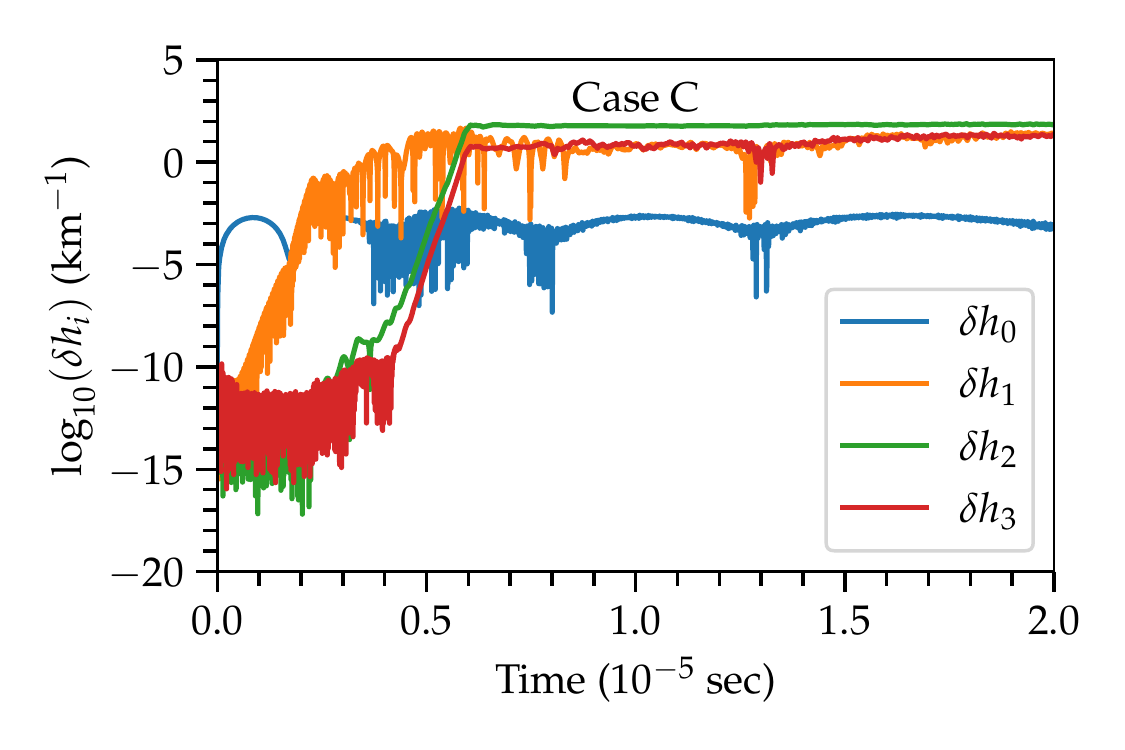}
\includegraphics[width=0.49\textwidth]{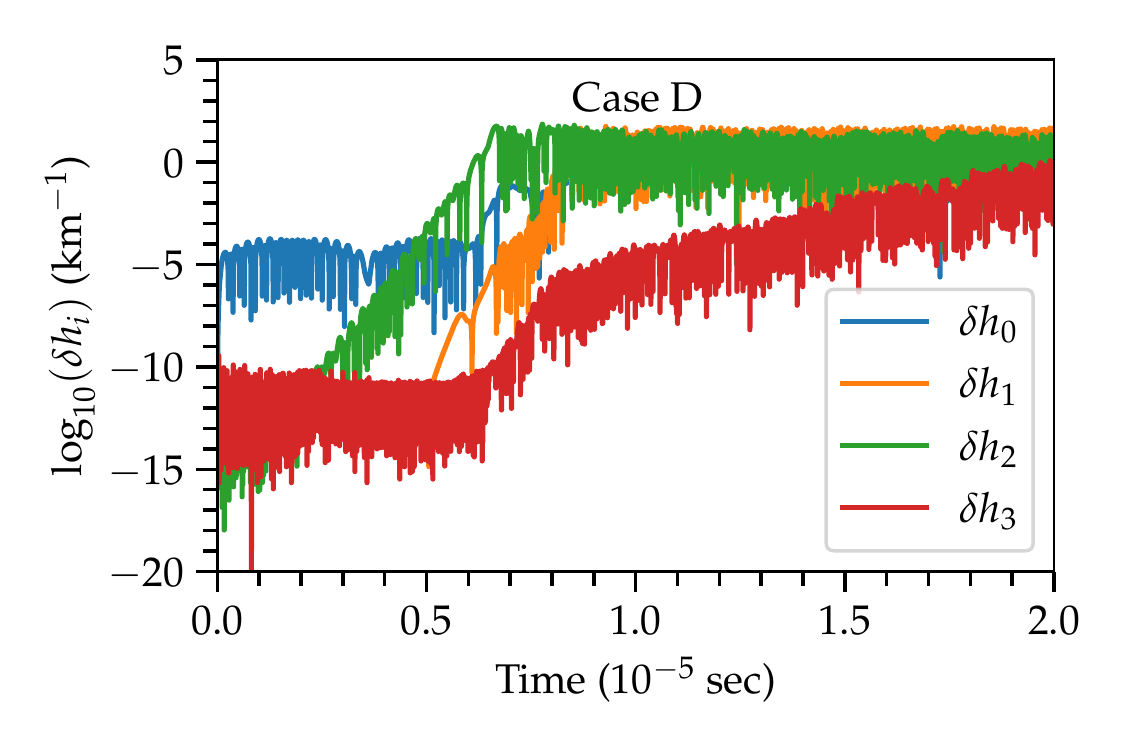}\\
\caption{Temporal evolution of various components of the Hamiltonian as defined in Eq.~\ref{eq14} for Cases A--D from top left  to bottom right, respectively.  The $y$-axis shows the deviation of the components of the Hamiltonian from their initial value at $t=0$. For Case A (B),  $h_{3}$ ($h_{1}$) is the dominant component initially;  for Case C (D), $h_{0}$ dominates at the initial time, however $h_{1}$ ($h_{2}$)  evolves the most during the initial times quickly reaching the non-linear regime.}
\label{FigS4}
\end{figure*}
In all cases, the reflection symmetry around $\phi=\pi$ is  broken  (see also the \href{https://sid.erda.dk/share_redirect/hwp0WKuuk7/index.html}{animations}). For Cases A and B,  the reflection symmetry is broken early in the linear regime, but the exponential growth is halted before it reaches magnitudes of order unity. For Case C, although the reflection symmetry is maintained in the initial stages of the non-linear regime, the symmetry is slowly broken on a small angular scale initially; then, it spreads to all angular scales, as visible from Fig.~\ref{Fig5}.  Case D  is qualitatively very different in that, as soon as the non-linear regime begins, there is a dramatic breaking of the reflection symmetry   and the exponential growth of the difference between $P_{ee}^{\alpha}$ and $P_{ee}^{\beta}$ continues to grow until it reaches magnitudes of order unity.

The evolution of the various components of the Hamiltonian defined as in Eq.~\ref{eq14} and displayed in Fig.~\ref{FigS4} reinforces the crucial role of non-linearity in the symmetry breaking, and the fact that linear stability analysis cannot give us an indication of the symmetries of the region along which the onset of flavor evolution occurs. For example,  the evolution of $H_{\nu\nu}$ for Case D shows that the most dominant component of the Hamiltonian is the one which is independent of the angle ($h_0$) at first, but the onset of flavor evolution occurs predominant along a circular region defined by $\phi=0, \pi$ (see also the \href{https://sid.erda.dk/share_redirect/hwp0WKuuk7/index.html}{animations}). The circular region where the onset of flavor mixing is located has the same symmetries as the $h_{3}$ term of the Hamiltonian. 

We have performed a calculation with several different angular distributions (results not shown here) and there are examples in which $h_{3}$ dominates  the onset of the nonlinear regime even though the $h_{2}$ term is the largest at the initial time. The symmetry or lack thereof is thus determined by a non-trivial combination of growth rates  in the linear regime and the initial values of various terms of the Hamiltonian, together with the initial ELN configuration and the conservation laws fulfilled by our system.

These findings are not an artifact of the three flavor calculations, but are exclusively dependent on the initial configurations of the system. The same trend is visible in the  two flavor scenario (see the \href{https://sid.erda.dk/share_redirect/hwp0WKuuk7/index.html}{animations});  the differences between the two and three flavor scenarios that we find for all cases including the polar and azimuthal angular variables are comparable to the ones discussed in Ref.~\cite{Shalgar:2021wlj}.  

\bibliography{thetaphi.bib}
\end{document}